\documentclass[preprint]{revtex4}
\usepackage{amsfonts}
\usepackage{amsmath}
\usepackage{amssymb}
\usepackage{graphicx}
\usepackage{booktabs}
\providecommand{\U}[1]{\protect\rule{.1in}{.1in}}

\begin{document}

\title{
{Surface Geometry of Some Meaningful Extremal Kerr-Newman Black Holes}
}

\author{Giorgio SONNINO}
\affiliation{
Universit{\'e} Libre de Bruxelles (U.L.B.)\\
Department of Physics\\
Campus de la Plaine C.P. 224 - Bvd du Triomphe\\
1050 Brussels - Belgium\\
Email: giorgio.sonnino@ulb.be
}

\affiliation{}


\begin{abstract}
\noindent 
\vskip 0.5truecm
We address the properties of extremal black holes by considering the Christodoulou-Ruffini/Hawking mass-energy formula. By simple geometrical arguments, we found that the mass/energy formula is satisfied by two meaningful extremal black holes where \textit{mass} ($m$), \textit{charge} ($Q$), and \textit{angular momentum/spin} ($L$) are proportional to the black holes \textit{irreducible mass} ($m_{ir}$) expressed by irrational numbers. These black holes have been studied in the Christodoulou diagram and their topology in $E^3$ has been investigated by differential geometry. We show that one of the analyzed Kerr-Newman black holes corresponds to the case where the Gaussian curvature becomes zero at the poles. in the first case, the coefficients are linked to irrational numbers. In the second extremal black hole examined, the fundamental quantities $m$, $Q$, and $L$ are linked to the irreducible mass by coefficients that depend solely on the \textit{golden ratio number} $-\phi_-$. In this case, we show that if this extremal black hole satisfies the Pythagorean fundamental forms relation at the umbilic points, then both the \textit{scale parameter} (corresponding to twice the irreducible mass) and the Gauss curvature of the surface at the poles are equal to the golden ratio numbers. For these two extremal black holes, we calculate the energy extractible by reversible transformations finding that, in percentage, the energy extractable from the latter black hole is higher than the former one.


\end{abstract} 

\maketitle


\section{Introduction}\label{I}
\noindent The Kerr-Newman black hole is a solution to the equations of general relativity that describe a rotating and charged black hole. It incorporates both the effects of angular momentum (spin) and electric charge, making it a more complex and realistic model than the non-rotating Schwarzschild black hole or the Reissner-Nordström black hole (which has charge but no spin). While real black holes are not expected to carry large electric charges, the study of charged black holes can provide insights into cosmological scenarios where charged black holes may have formed in the early universe. These scenarios could involve interactions with electrically charged particles during the universe's early stages. Many astrophysical objects, such as accreting black holes in binary systems or active galactic nuclei, may possess both angular momentum and electric charge. The Kerr-Newman solution helps physicists and astronomers understand the effects of rotation and charge on the behavior of these objects, including how they emit radiation, affect their surroundings, and influence their accretion processes. Additionally, studying the Kerr-Newman solution provides opportunities to test the predictions of general relativity under more complex conditions. The curvature of spacetime around a rotating charged black hole is different from that of simpler black holes, and studying these differences allows for tests of the theory's predictions in extreme scenarios. Investigating the extreme Kerr-Newman black hole is important for gaining a deeper understanding of the behavior of rotating, charged black holes and their relevance to astrophysical and theoretical contexts. It enables researchers to test the predictions of general relativity, explore the relationships between gravity and thermodynamics, and contribute to our broader understanding of fundamental physics. The description of black holes is notably simplified by adopting the validity of the so-called \textit{No-Hair Theorem} \cite{wheeler, carter}. The No-Hair Theorem is a concept in theoretical physics that suggests that the observable properties of a black hole, such as its mass, charge, and angular momentum (spin), are the only relevant characteristics that can be determined from the outside. In other words, the complex details of the matter that formed the black hole are not important for describing its long-term behavior. Studies involving black hole mergers and gravitational wave detections have provided further support for the no-hair theorem in recent years. Nonetheless, exploring black holes and their properties remains an active area of research, and our understanding of these enigmatic objects continues to evolve. Recent investigations showed that studying extreme black holes may shed some light on the black holes' feature properties outside of the three classical black hole traits of mass, spin, and charge \cite{burko}. This paper investigates two significant extremal-charged rotating black holes that can easily be detected simply by looking at the geometric version of the mass/energy formula for extreme Kerr-Newman black holes. The topology of event horizon surfaces will be studied using the tools provided by differential geometry. When a black hole rotates, its surface deforms; there is the characteristic flattening of the poles and lengthening of the equatorial circumference. The first case corresponds to the extreme Kerr-Newman black hole where the Gaussian curvature of the surface starts to flatten at the poles. The second case is much more spectacular and intriguing. We prove that if this extreme black hole satisfies the Pythagorean fundamental forms relation at the umbilic points, all fundamental quantities (mass, charge, and angular momentum) are equal to the square root of the golden section raised to odd integers. Additionally, both the \textit{scale parameter} and the Gauss curvature of the surface at the pole are equal to the golden ratio numbers. For this special extreme black hole, the mean curvature of the black hole surface is equal to the square root of the golden ratio. However, since the extrinsic curvature depends on the embedding space our analysis will be mainly focused on the intrinsic geometry.

\noindent The work is organized as follows. In section~\ref{knbh} we describe extreme black holes in the Kerr-Newman geometry. Section~\ref {MEBH} shows how two meaningful extreme Kerr-Newman black holes can easily be detected simply by looking at the geometrical version of the mass/energy extreme Kerr-Newman black holes. The location and description of the two extreme black holes in the Christodoulou diagram can be found in section~\ref{christodoulou}. The topology of these two extreme black holes is studied by differential geometry in section~\ref{GBH}. In section~\ref{er} we compute the energy extractable from these extreme Kerr-Newman black holes by reversible transformations. Concluding remarks can be found in section~\ref{c}. The proofs of the theorems are reported in the Appendices.

\section{Extremal Black Holes in the Kerr-Newman Geometry}\label{knbh}
\noindent In the Kerr-Newman geometry, the horizon is located at \cite{wheeler}
\begin{equation}\label{s1}
r_{+}=m+\sqrt{m^2-Q^2-a^2}
\end{equation}
\noindent where $m$ is the mass of the black hole, $Q$ is the charge, and $a =L/m$ is the angular momentum per unit mass. In Eq.~(\ref{s1}) we used \textit{Planck's unit system} where the five universal physical constants i.e., the \textit{speed of light} $c$, the \textit{Universal gravitational constant} $G$, the \textit{reduced Planck's constant} $\hbar$, \textit{Boltzmann's constant} $k_B$, and the \textit{vacuum permittivity} $\varepsilon_0$ take on the numerical value of $1$ when expressed in terms of these units \footnote{The modern values for Planck's original choice of quantities are: \textit{Planck's length} = $\sqrt{\hbar G/c^3}$, \textit{Planck's mass} = $\sqrt{\hbar c/G}$, \textit{Planck's angular momentum} = $\hbar$, \textit{Planck's temperature} = $\sqrt{\hbar c^5/(G k_B^2)}$, \textit{Planck's energy} = $\sqrt{\hbar c^5/G}$, and \textit{Planck's charge} = $\sqrt{4\pi\varepsilon_0 \hbar c}=e/\sqrt\alpha$ with $e$ and $\alpha$ denoting the electric charge and the fine-structure constant, respectively.}. Unless otherwise indicated, Planck's unit system will be adopted also in the sequel. The Kerr-Newman black hole exists if 
\begin{equation}\label{s2}
m^2\geq Q^2+a^2
\end{equation}
\noindent Any collapsing body that violates this constraint, centrifugal forces and/or electrostatic repulsion will halt the collapse before a size $\sim m$ is reached. Notice that due to the no-hair theorem, all information about the matter in the hole is missing for an outside observer except the total mass, angular momentum, and charge of the black hole. The general equation (\ref{s1}) can be rewritten in an alternative form by expressing the mass-energy formula of a Kerr-Newman black hole $m$ in terms of the irreducible mass $m_{ir}$ (as well as a function of the charge $Q$ and of the angular momentum $L$)  \cite{christodoulou, hawking}:
\begin{equation}\label{s3}
m^2=\left(m_{ir}+\frac{Q^2}{4m_{ir}}\right)^2+\frac{L^2}{4m_{ir}^2}
\end{equation}
\noindent where the irreducible mass is defined as
\begin{equation}\label{s4}
m_{ir}\equiv\frac{1}{2}\sqrt{r_+^2+a^2}
\end{equation}
\noindent In this form, the Kerr-Newman black hole exists if \cite{ruffini} 
\begin{equation}\label{s5}
\frac{L^2}{4m_{ir}^4}+\frac{Q^4}{16m_{ir}^4}\leq1
\end{equation}
\noindent The surface area of the event $A$ (i.e., the horizon) is located at $r_h=$\textit{gravitational radius} \cite{wheeler}. So
\begin{equation}\label{s6}
A=4\pi r_h^2
\end{equation}
\noindent Only the region on and outside the black hole's surface, $r\geq 2m_{ir}$, is relevant to external observers \cite{ruffini}. Events inside the horizon can never influence the exterior. The horizon is related to the irreducible mass according to the equation $r_h=2m_{ir}$ and the expression of the surface area of the event reads \cite{ruffini}
\begin{equation}\label{s7}
A=16\pi m_{ir}^2\quad{\rm so}\quad\frac{A}{4\pi}=r_+^2+a^2=2m(m+\sqrt{m^2-Q^2-a^2})-Q^2
\end{equation}
\vskip0.5truecm
\noindent $\bullet$ {\bf The extreme Kerr-Newman Black Holes}

\noindent The extremal black hole is described by the equations
\begin{align}\label{s9}
&m^2=\left(m_{ir}+\frac{Q^2}{4m_{ir}}\right)^2+\frac{L^2}{4m_{ir}^2}\\
&\frac{L^2}{4m_{ir}^4}+\frac{Q^4}{16m_{ir}^4}=1\nonumber
\end{align}
\noindent Eqs~(\ref{s6}) can be brought in the following form
\begin{align}\label{s9}
&(\sqrt{2}m_{ir})^2+\left(\frac{Q}{\sqrt{2}}\right)^2=m^2\\
&a^2+\left(\frac{Q}{\sqrt{2}}\right)^2=(\sqrt{2}m_{ir})^2\nonumber
\end{align}
\noindent Eqs~(\ref{s9}) are the equations for an extreme Kerr-Newman black hole. Note that substituting the expression for $2m_{ir}$ given by the first equation of system~(\ref{s9}) into the second equation of system~(\ref{s9}) we re-obtain
\begin{equation}\label{s10}
a^2+Q^2=m^2
\end{equation}
\noindent Substituting the expression for $Q/\sqrt{2}$ given by the first equation of system~(\ref{s9}) into the second equation of system~(\ref{s9}) we get
\begin{equation}\label{s11}
a^2+m^2=(2m_{ir})^2
\end{equation}
\noindent So, $m_{ir}\leq m\leq 2m_{ir}$ and the minimum value of energy $m$ corresponds to a Schwarzschild black hole ($L=0$ and $Q=0$) while its maximum value to the extreme Reissner–Nordström black hole ($L=0$). It is convenient to get rid of numbers like $2$ or $\sqrt{2}$ and rewrite the equations for the extreme Kerr-Newman black holes in terms of the following \textit{scaled quantities}:
\begin{align}\label{s12}
&{\bar L}\equiv 2L\\
&{\bar m}\equiv \sqrt{2}m \nonumber\\
&\eta\equiv 2 m_{ir}\nonumber
\end{align}
\noindent In the literature, parameter $\eta$ is called the \textit{scale parameter} \cite{smarr2}. In terms of variables ${\bar L}$, $\bar m$, and $\eta$, the equations for the extreme Kerr-Newman black holes notably simplify:
\begin{align}\label{s13}
&\eta^2+Q^2={\bar m}^2\\
&{\bar a}^2+Q^2=\eta^2\nonumber
\end{align}
\noindent where ${\bar a}={\bar L}/{\bar m}=\sqrt{2}\ \!a$. In the Appendix, we can check that Eqs~(\ref{s9}), (\ref{s10}), and (\ref{s11}) are equivalent, from a geometrical point of view to four right triangles interconnected each with the other (see in Annex Fig.~\ref{fig_EBH1}). 

\section{Two Meaningful Extremal Kerr-Newman Black Holes}\label{MEBH}
\noindent Two very special extremal black holes deserve attention. We denote these as case~(i) and case~ (ii), satisfying the following special conditions:
\subsection{Case (i)}
\begin{equation}\label{c5}
|Q|=\frac{1}{\sqrt{2}}\eta
\end{equation}
\noindent In this case, from system~(\ref{s9}) we get 
\begin{align}\label{c6}
&{\bar m}=\sqrt{\frac{3}{2}}\eta\\
&|Q|=\frac{1}{\sqrt{2}}\eta\nonumber\\
&|{\bar L}|=\frac{\sqrt{3}}{2}\eta^2 \nonumber
\end{align}
\noindent From the geometrical point of view, Eq.~(\ref{c5}) corresponds to the particular configuration where the triangle ABC coincides with triangle ACE (see in Appendix Fig.~\ref{fig_EBH5}).

\subsection{Case (ii)}
\begin{equation}\label{c1}
{|\bar L|}=\eta |Q|
\end{equation}
\noindent From the geometrical point of view, Eq.~(\ref{c1}) corresponds to the particular configuration where the catheter BD of the right triangle ADB coincides exactly with the height of the triangle ABC relative to the base AC (see in Appendix Fig.~\ref{fig_EBH2}). It is curious to see that in this case the charge $|Q|=|L|/m_{ir}$ turns out to be a combination of the angular momentum $L$ with the irreducible mass $m_{ir}$, which is an intrinsic quantity of the Kerr-Newman metric. By plugging condition~(\ref{c1}) into Eqs~(\ref{s13}) we find
\begin{align}\label{c2}
&{\bar m}=(-\phi_-)^{-1/2}\eta\\
&|Q|=(-\phi_-)^{1/2}\eta\nonumber\\
&|{\bar L}|=(-\phi_-)^{1/2}\eta^2 \nonumber
\end{align}
\noindent where $\phi_-$ is one of the solutions of the golden ratio equation
\begin{equation}\label{c3}
\phi^2-\phi-1=0
\end{equation}
\noindent i.e.,
\begin{equation}\label{c4}
\phi_{-}= \frac{1-\sqrt{5}}{2}
\end{equation}
\noindent Note that the other solution $\phi_+$ of the golden ratio equation~(\ref{c3}) is linked to $\phi_-$ by the relation $\phi_+=(-\phi_-)^{-1}$. Eq.~(\ref{c2}) states that \textit{there exists a whole family of charged rotating extremal black holes in which the fundamental quantities are incommensurate with its irreducible mass and the irrational constants depend solely on the golden ratio}. In the next section, we shall analyze the extreme black holes~(\ref{c6}) and (\ref{c2}) in Christodoulou's diagram.  

\section{The Two Extremal Kerr-Newman Black-Holes in Christodoulou Diagram}\label{christodoulou}

\subsection{ The Christodoulou diagram}
\begin{figure}
\includegraphics[width=4cm]{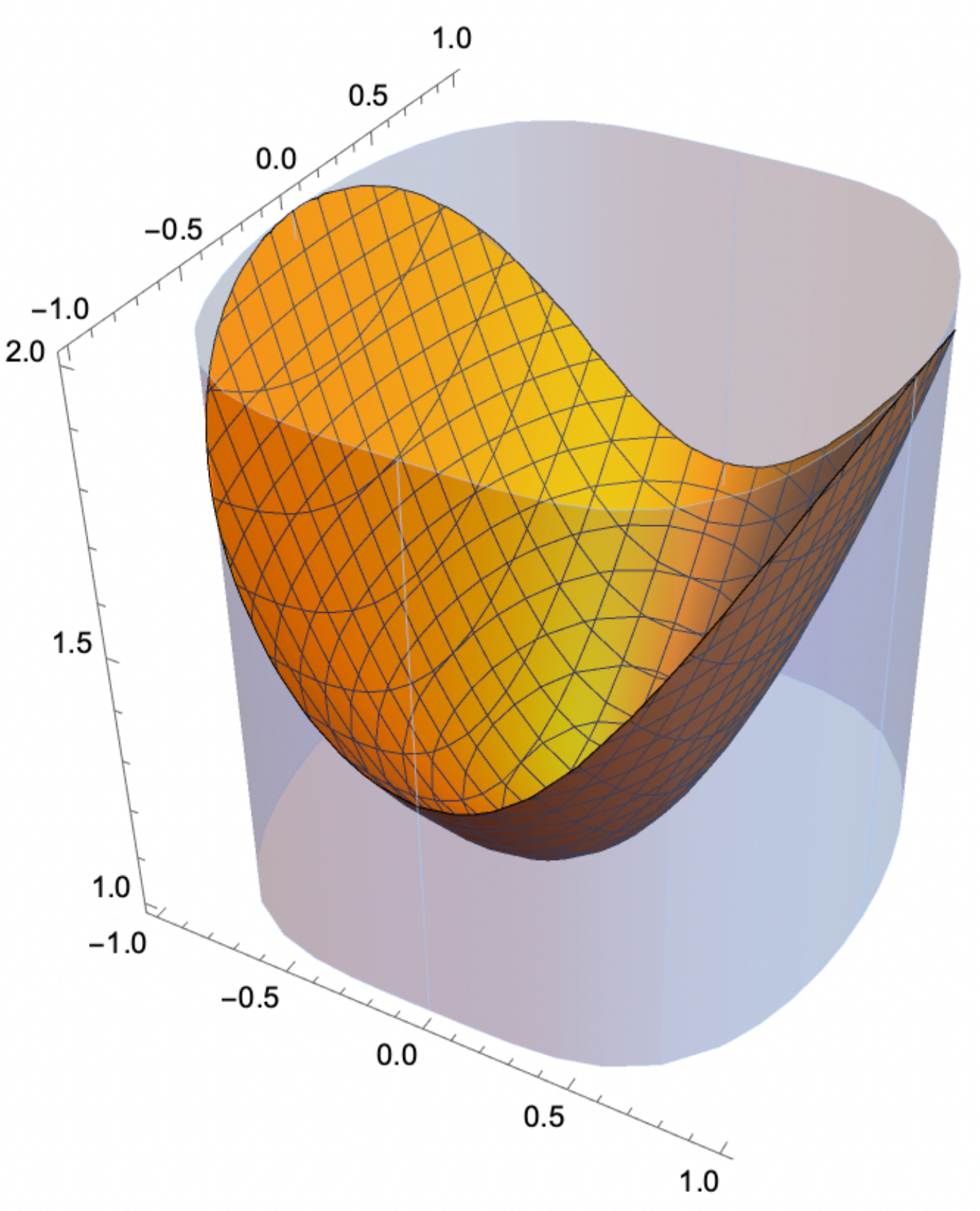}
\caption {\textit{Mass/Energy formula for the Kerr-Newman black holes (the orange surface). These black holes exist only in the range $x^4+y^2\leq 1$ (the grey zone).}}
\label{fig_EBH6}
\end{figure}

\noindent Christodoulou's diagram plots the contours $m/m_{ir}$ in the plane scaled angular momentum $L/m_{ir}^2$ versus scaled charge $Q/m_{ir}$ \cite{christodoulou1}. Black holes exist only in the interior of the region $a^2+Q^2\leq m^2$. The Christodoulou diagram is easily obtained by introducing the following scaled variables
\begin{align}\label{a05}
&x=\frac{Q}{\eta}\\
&y=\frac{{\bar L}}{\eta^2}\nonumber\\
&z=\sqrt{2} \frac{{\bar m}}{\eta}=\frac{m}{m_{ir}}\nonumber
\end{align}
\noindent In the new variables, the equations for the Kerr-Newman black hole (\ref{s3}) and (\ref{s5}) read
\begin{equation}\label{a06}
z^2-\left(1+x^2\right)^2-y^2=0\qquad{\rm with}\qquad x^4+y^2\leq 1
\end{equation}
\noindent The extreme black holes satisfy the equation $x^4+y^2=1$. Accordingly, one finds the range of $x$, $y$, and $z$  for the family of black holes associated with the given $\eta$:
\begin{equation}\label{a07}
-1\leq x\leq 1\quad ;\quad -1\leq y\leq 1\quad {\rm and}\quad 1\leq z\leq 2
\end{equation}
\noindent Fig.~\ref{fig_EBH6} shows the behavior of the mass/energy formula for a Ker-Newman black hole. These black holes exist in the ranges~(\ref{a07}).
\noindent It is easily checked that equations~(\ref{c6}) and (\ref{c2}) refer to extreme black holes. Indeed, for the extreme black hole~(\ref{c6}), with a positive charge and angular momentum, we get
 \begin{align}\label{a08}
&x=\frac{|Q|}{\eta}=\frac{1}{\sqrt{2}}=0.707107\\
&y=\frac{|{\bar L}|}{\eta^2}=\frac{\sqrt{3}}{2}=0.866025\quad {\rm so}\nonumber\\
&x^4+y^2=\frac{1}{4}+\frac{3}{4}=1\nonumber
\end{align}
 \noindent in agreement with Eq.~(\ref{s10}). For the extreme black hole~(\ref{c2}), with a positive charge and angular momentum, we find
 \begin{align}\label{a09}
&x=\frac{|Q|}{\eta}=(-\phi_-)^{1/2}=\frac{\sqrt{\sqrt{5}-1}}{\sqrt{2}}=0.786151\\
&y=\frac{|{\bar L}|}{\eta^2}=(-\phi_-)^{1/2}=\frac{\sqrt{\sqrt{5}-1}}{\sqrt{2}}=0.786151\qquad{\rm so}\nonumber\\
&x^4+y^2=\phi_-^2-\phi_-=1+\phi_--\phi_-=1\nonumber
\end{align}
 \noindent where we have taken into account the identity $\phi_-^2=1+\phi_-$.

\subsection{Location of the extremal black holes~(\ref{c6}) and (\ref{c2}) in Christodoulou's diagram}
\begin{figure}
\includegraphics[width=6cm]{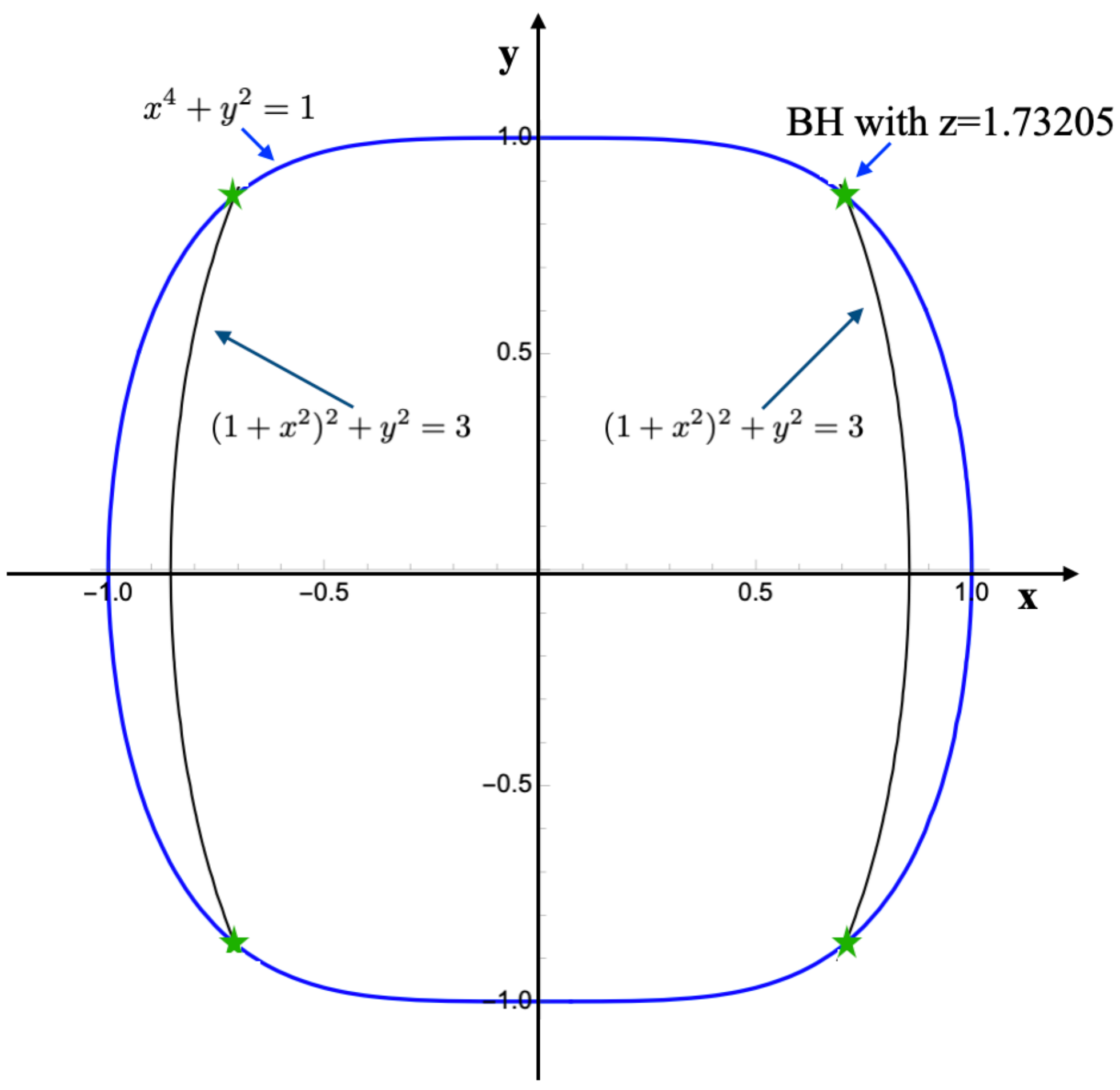}
\caption {\textit{Scaled Christodoulou's diagram. Contours of constant $z=m/m_{ir}$ are depicted in the x-y plane. Black holes can exist only in the interior of the region $x^4+y^2\leq 1$. Black holes in the blue contour are the extreme black holes. The green stars correspond to the extreme black holes~(\ref{c6}).}}
\label{fig_EBH7a}
\end{figure}

\begin{figure}
\includegraphics[width=6cm]{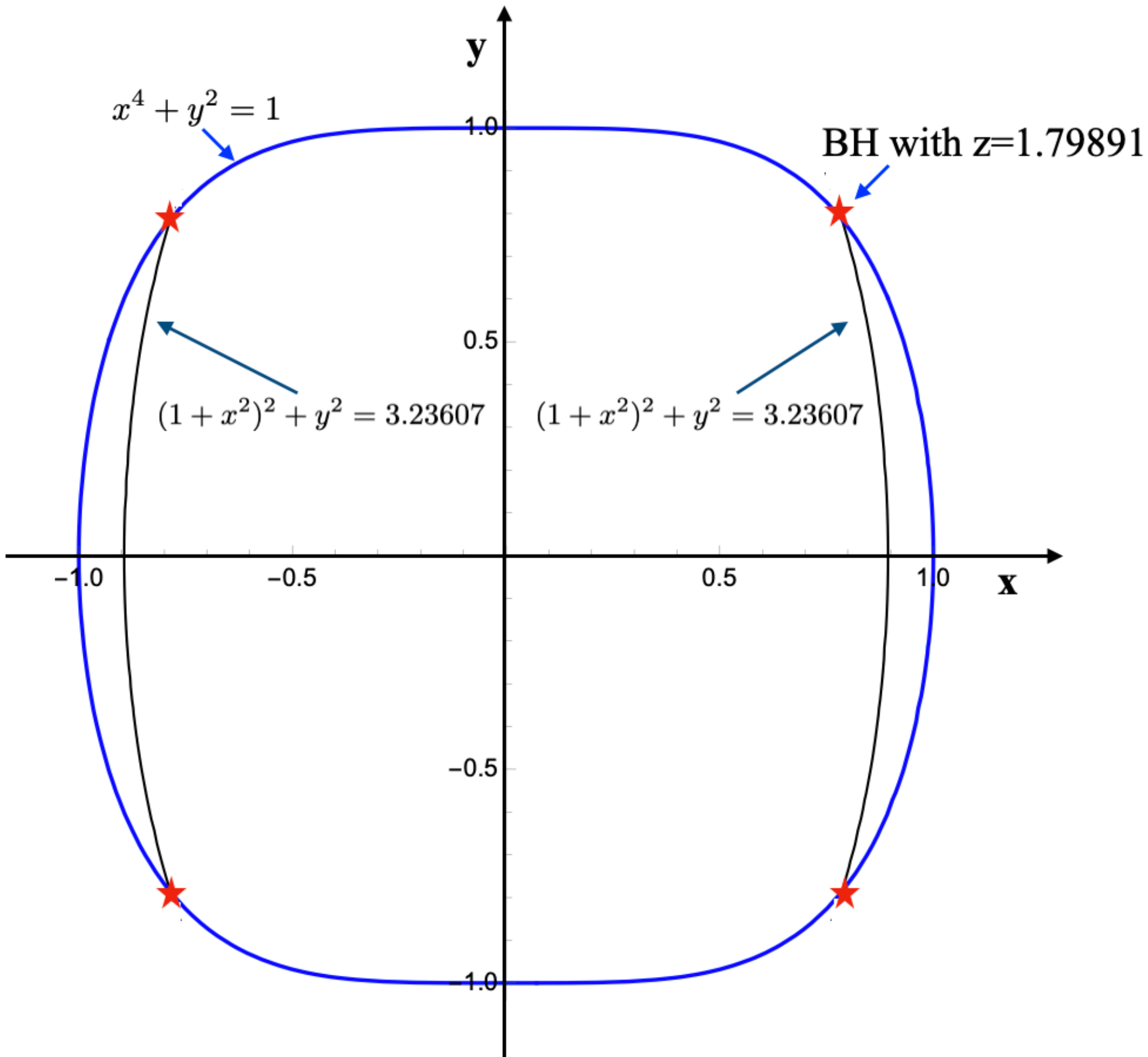}
\caption {\textit{Scaled Christodoulou's diagram. Contours of constant $z=m/m_{ir}$ are depicted in the x-y plane. The extreme black holes are located in the blue curve $x^4+y^2\leq 1$. The red stars refer to the extreme black holes~(\ref{c2}), Note that black holes marked by red stars are the only extreme Kerr-Newman black holes that satisfy the symmetry $|y|\leftrightarrow |x|$.}}
\label{fig_EBH7b}
\end{figure}

\begin{figure}
\includegraphics[width=7cm]{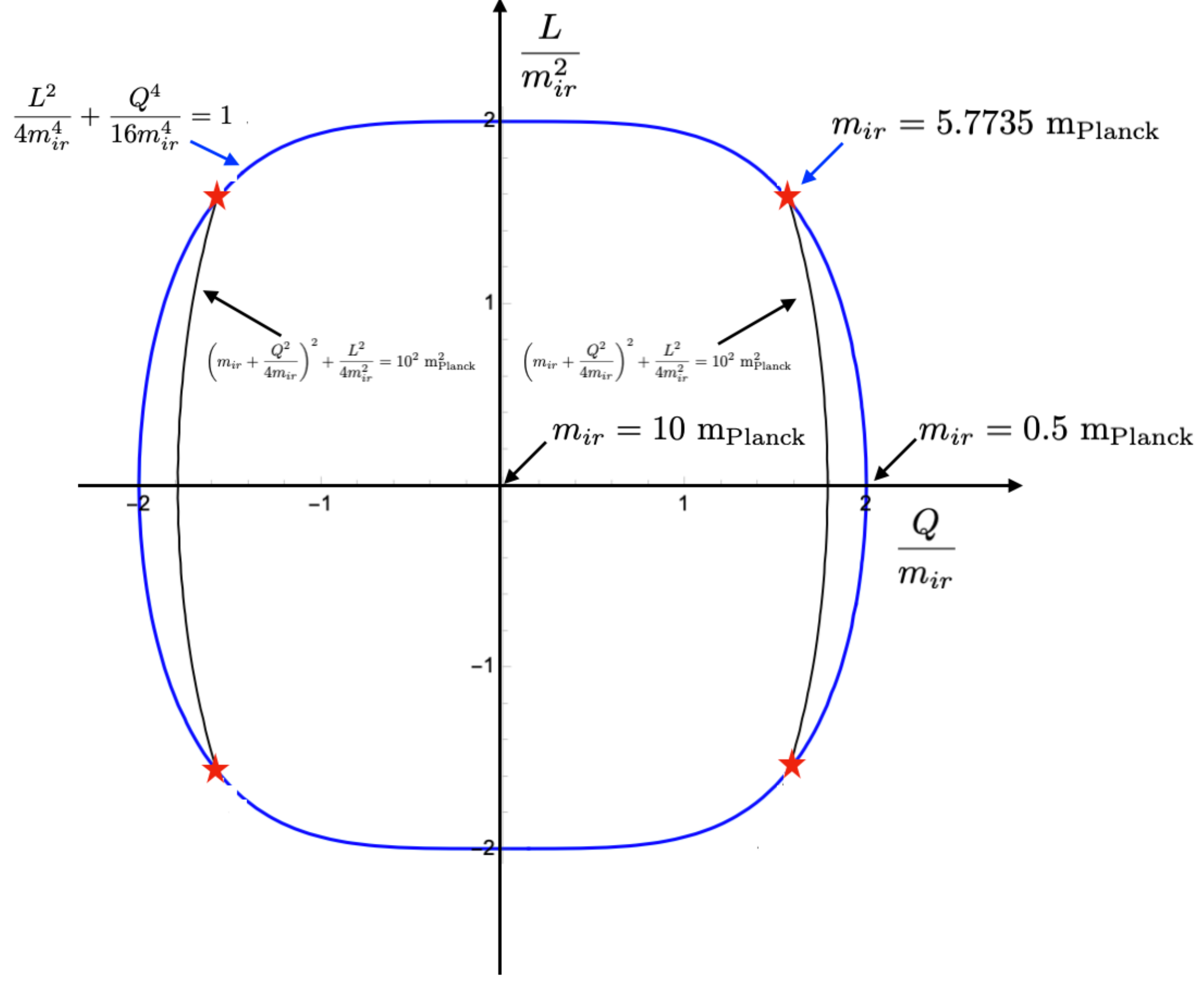}
\caption {\textit{Dimensional Christodoulou's diagram. The figure shows the extreme black hole with mass ${\rm m=10 m_{Planck}}$ and satisfying the symmetry $|y|\leftrightarrow |x|$. The blue contour corresponds to the extremal black holes satisfying the relation $4L^2+Q^4=16m_{ir}^4$. The black contours are the mass/energy formula with $m_{ir}=5.7735\ m_{Planck}$.}}
\label{fig_EBH7c}
\end{figure}

\begin{figure}
\includegraphics[width=7cm]{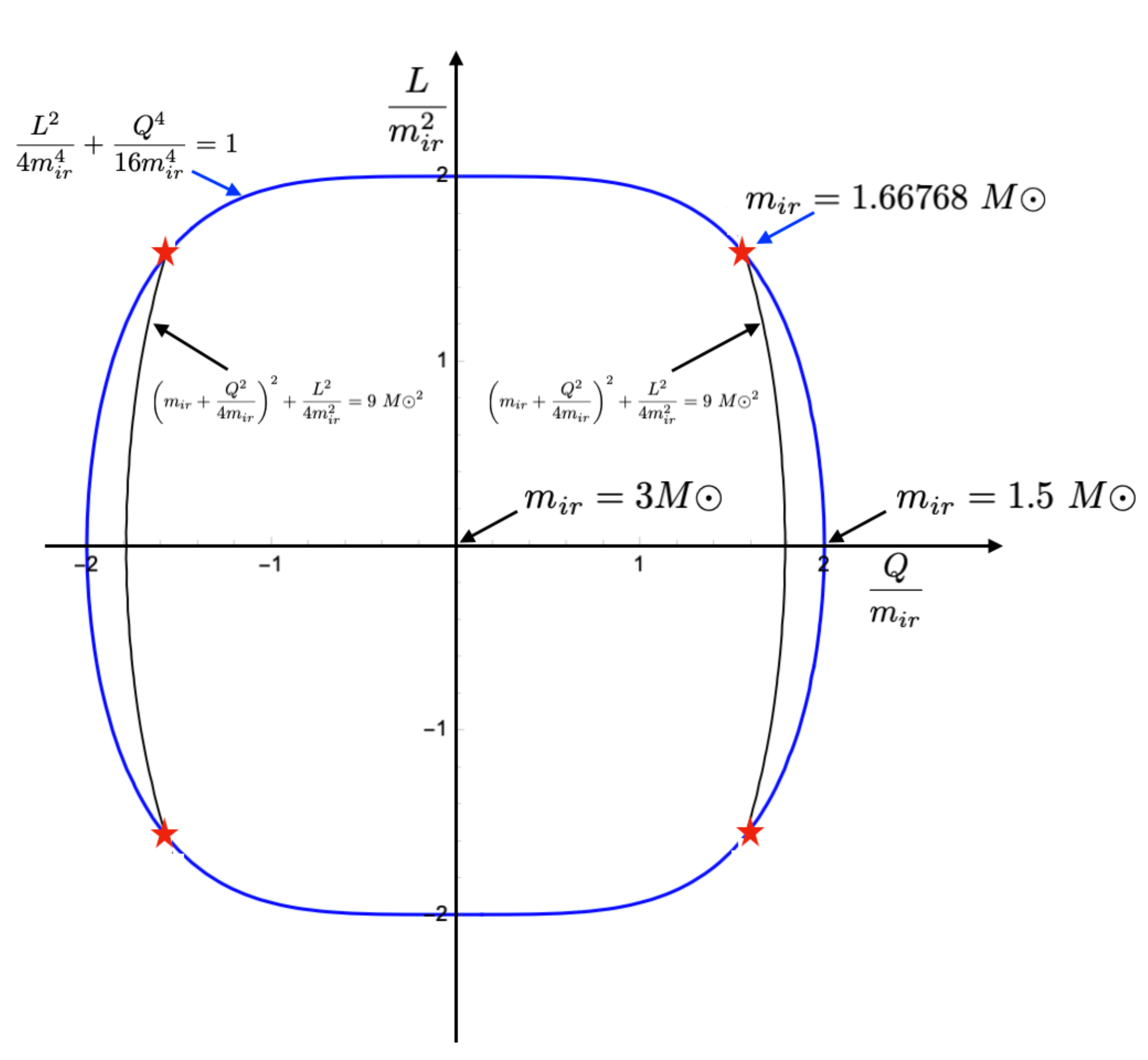}
\caption {\textit{Dimensional Christodoulou's diagram. The figure shows the extreme black hole with mass $m=3M\odot$ and satisfying the symmetry $|y|\leftrightarrow |x|$. The black contours are the mass/energy formula with $m_{ir}=1.66768M\odot$.}}
\label{fig_EBH7d}
\end{figure}

\begin{figure}
\includegraphics[width=7cm]{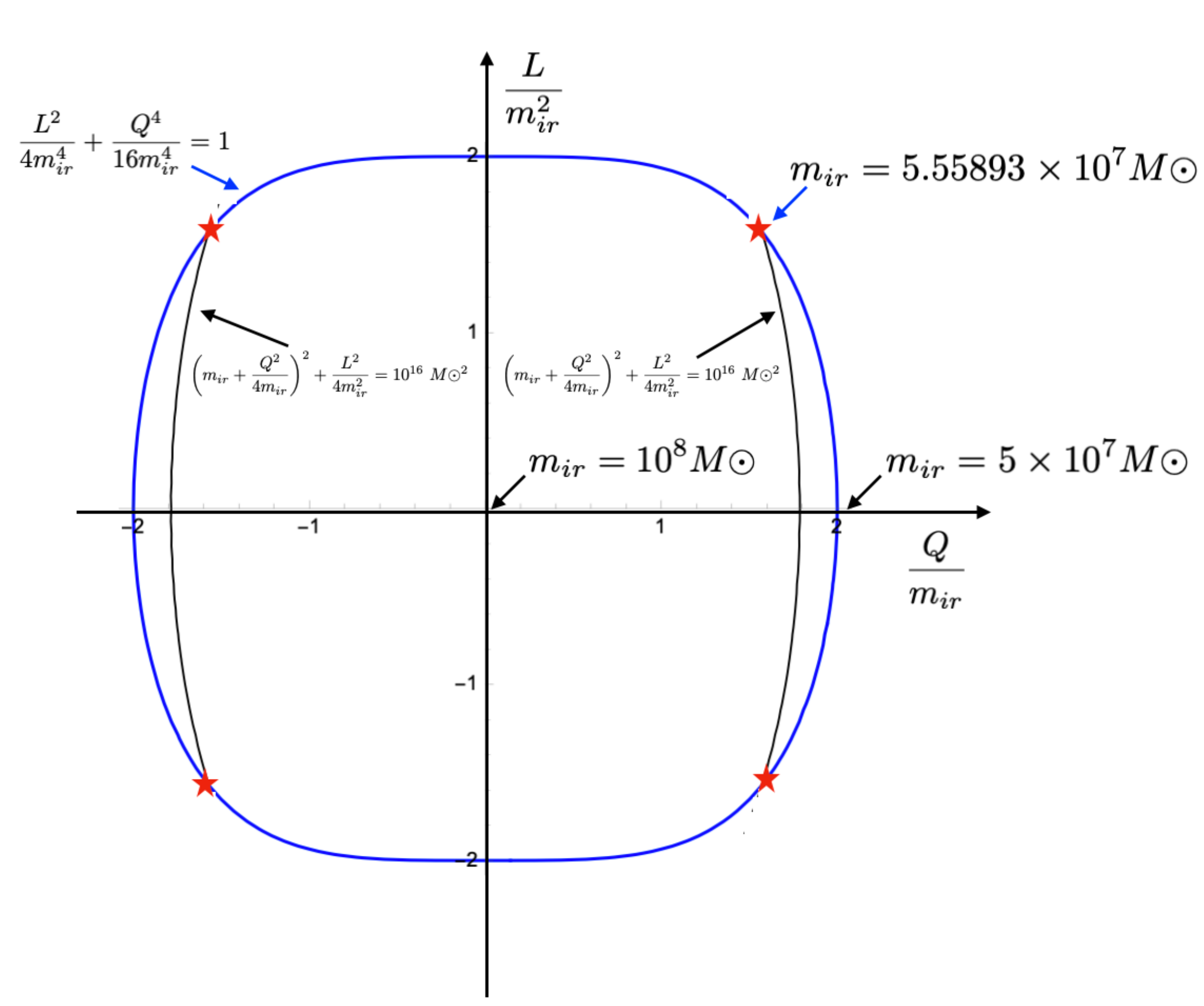}
\caption {\textit{Dimensional Christodoulou's diagram. The figure shows the extreme black hole with mass $m=10^8M\odot$ and satisfying the symmetry $|y|\leftrightarrow |x|$. The black contours are the mass/energy formula with $m_{ir}=5.55893\times 10^7M\odot$.}}
\label{fig_EBH7e}
\end{figure}

\noindent Case {\bf (i)}

\begin{align}\label{a010}
& {\rm Contour}:\ z_c=\sqrt{3}=1.73205\\
& {\rm Coordinate\ of\ the\ black \ hole}: \left\{x, y \right\}=\left\{1/\sqrt{2}, \sqrt{3}/2 \right\}\nonumber
\end{align}
\noindent Case {\bf (ii)}
\begin{align}\label{a011}
& {\rm Contour}:\ z_c=\frac{\sqrt{2}}{\sqrt{-\phi_-}}=1.79891\\
& {\rm Coordinate\ of\ the\ black \ hole}: \left\{x, y\right\}=\left\{(-\phi_-)^{1/2}, (-\phi_-)^{1/2}\right\}\nonumber
\end{align}
\noindent Fig.~\ref{fig_EBH7a} and Fig.~\ref{fig_EBH7b} show the location of the two extreme black holes~(\ref{c6}) and  (\ref{c2}), respectively, in the \textit{scaled Christodoulou diagram}.
\noindent As an example, Fig.~\ref{fig_EBH7c}, Fig.~\ref{fig_EBH7d}, and Fig.~\ref{fig_EBH7e} show three black holes~(\ref{c2}) in the \textit{dimensional Christodoulou's diagram}. These black holes have mass values that differ from each other by several orders of magnitude, ranging from a microscopic value ($m=10\ m_{Planck}$, see Fig.~\ref{fig_EBH7c}) to a macroscopic one ($m=10^8M\odot$, see Fig.~\ref{fig_EBH7d}), passing from an intermediate value ($m=3M\odot$, see Fig.~\ref{fig_EBH7e}). Here, $m_{Planck}$ and $M\odot$ denote Planck's mass (${\rm m_{Planck}}=2.18\times 10^{-8}{\rm Kg}$) and the sun mass ($M\odot\sim2\times 1030 {\rm Kg}$), respectively.

\noindent Notice that the coordinates of the other six extreme black holes, indicated by the green and red stars, can be easily obtained by setting $z_c=1.73205$, for the case {\bf (i)}, and $z_c=1.79891$, for the case {\bf (ii)}, in Eq.~(\ref{a06}), respectively, and solving the equations for variables $x$ and $y$.

\subsection{Symmetries}
\noindent As we can see, the plots of the Christodoulou diagram are symmetric concerning the following transformations $\{x,y\}\rightarrow\{-x,y\}$ i.e., symmetry concerning the reflection about the $L$-axis (the \textit{Kerr axis}), $\{x,y\}\rightarrow\{x,-y\}$ i.e., symmetry for the reflection about the $Q$-axis (the \textit{Reissner–Nordstr$\ddot{o}$m axis}), and $\{x,y\}\rightarrow\{-x,-y\}$ i.e., symmetry concerning the origin of the axes $(Q,L)=(0,0)$ (the \textit{Schwarzschild point}). We can easily check the black holes~(\ref{c2}) (indicated by red stars in this diagram) are the only ones that are invariant under the exchange of $|y|$ with $|x|$ (and vice-versa). Indeed, extreme Kerr-Newman black holes satisfy the equation
\begin{equation}\label{a012}
x^4+y^2-1=0
\end{equation}
\noindent We are looking for extreme black holes that are invariant concerning the exchange of $|y|$ with $|x|$, and vice-versa. This symmetry requires that $x$ satisfies the equation
\begin{equation}\label{a013}
x^4+x^2-1=0
\end{equation}
\noindent Eq.~(\ref{a013}) admits only one positive solution $x^2$ given by
\begin{equation}\label{a014}
x^2=-\phi_-\qquad{\rm or}\qquad |x|=(-\phi_-)^{1/2}=|y|
\end{equation}
\noindent Plugging these solutions into the mass/energy formula (\ref{a06}), we get the value of $z$
\begin{equation}\label{a015}
z=\sqrt{2\phi_+}=\frac{\sqrt{2}}{\sqrt{-\phi_-}}
\end{equation}
\noindent in agreement with Eqs~(\ref{a09}) and (\ref{a011}). In short, \textit{the extreme Kerr-Newman black holes marked by red stars are quite peculiar since they belong to the family of black holes located on the bisectors of the Christodoulou diagram}. The next section is devoted to the study of the geometrical properties of the two extreme black holes {\bf (i)} and {\bf (ii)}.

\section{Study of the Surface Geometry of the Extremal Black Holes}\label{GBH}
\begin{figure}
\includegraphics[width=4cm]{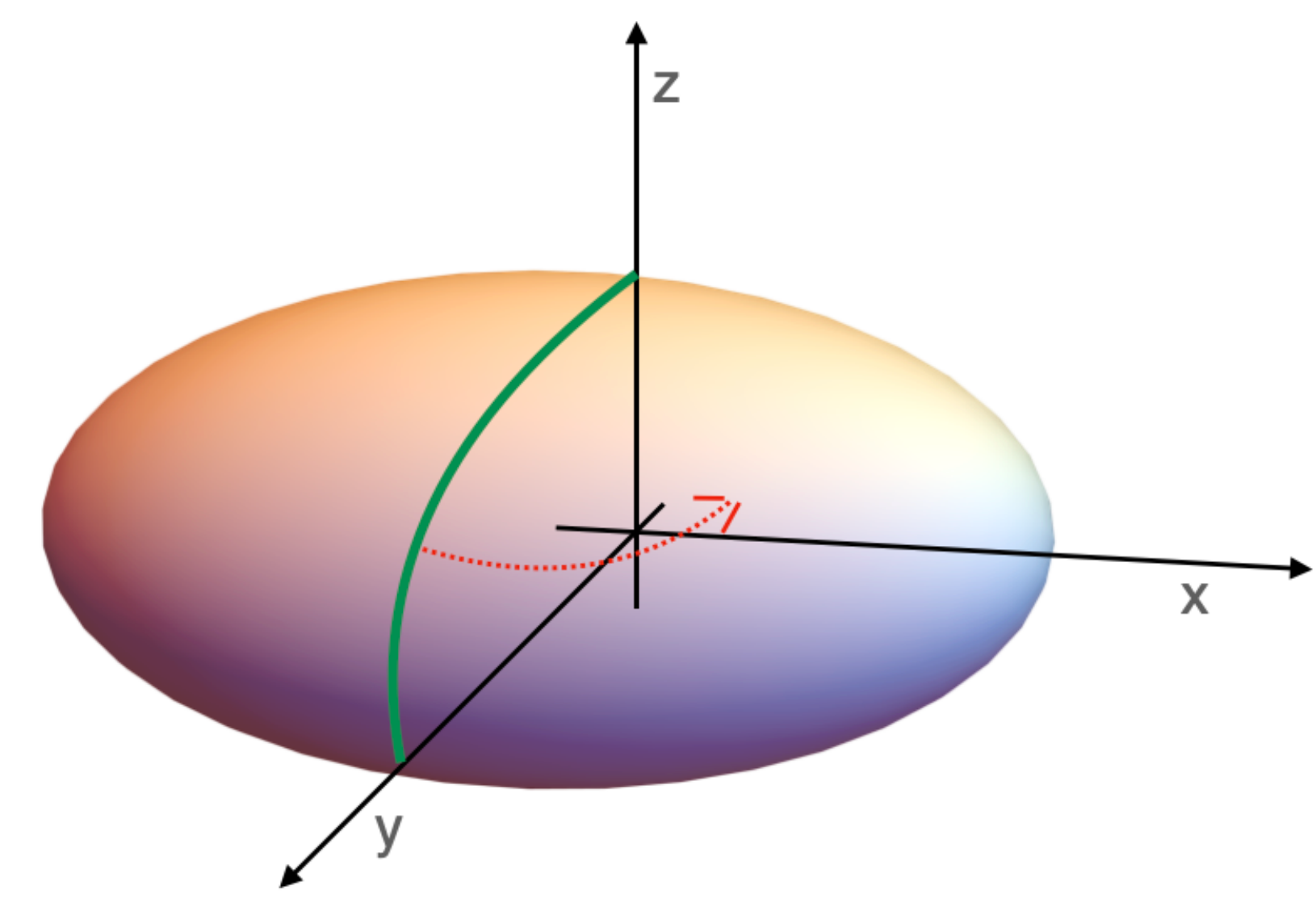}
\caption {\textit{The horizon surface of revolution of a black hole embedded in the Euclidean 3-space.}}
\label{fig_EBH3}
\end{figure}

\noindent To visualize the intrinsic geometry of a black hole we embed the surface isometrically in the Euclidean 3D space $E^3$ \cite{smarr2}. First, it is convenient to introduce a metric function $h(\mu)$ defined as 
\begin{equation}\label{gsi1}
h(\mu)=(1-\mu^2)[1-\beta^2(1-\mu^2)]^{-1}\qquad{\rm where}\quad \beta=\frac{a}{\eta}
\end{equation}
\noindent In the literature, parameter $\beta$ is called \textit{distortion parameter}. In Kerr-Newman geometry, the metric is written in a standard form \cite{carter}, \cite{smarr2}
\begin{equation}\label{gsi2}
ds^2=\eta^2(h(\mu)^{-1}d\mu^2+h(\mu)d\psi^2)\quad {\rm with}\quad -1\leq\mu\leq 1\ \ ,\ \ 0\leq\psi\leq 2\pi
\end{equation}
\noindent The values $\mu=\pm 1$ correspond to the poles. The equator corresponds to the value $\mu=0$. The procedure for investigating an arbitrary surface of revolution in $E^3$ is standard. As a curve ${\mathcal C}$ in the ${\rm xz}$-plane ${\rm x}=f(\nu)$, ${\rm z}=g(\nu)$
revolves about the ${\rm z}$-axis, it generates a surface of revolution ${\mathcal S}$. The curves ${\mathcal C}$ in different rotated positions are the meridians of  ${\mathcal S}$. In contrast, the circles generated by each point on ${\mathcal C}$ are the parallels of ${\mathcal S}$ (see Fig.~\ref{fig_EBH3}). 
\noindent If we denote the rotation angle in the ${\rm xy}$-plane as $\theta$, the surface of revolution can be parametrized as
\begin{equation}\label{gsi3}
({\rm x,y,z})=(f(\nu)\cos\theta,f(\nu)\sin\theta,g(\nu))
\end{equation} 
\noindent Equating the 2-metric resulting from~(\ref{gsi3})
\begin{equation}\label{gsi4}
ds^2=(f'{}^2+g'{}^2)d\nu^2+f^2d\theta^2
\end{equation} 
\noindent with metric~(\ref{gsi2}), we get $\nu=\mu$, $\theta=\psi$ and
\begin{equation}\label{gsi4}
f(\mu)=\eta h^{1/2}\qquad {\rm ;}\qquad g'(\mu)=h^{-1/2}\left(1-\frac{1}{4} {h'}^2\right)^{1/2}
\end{equation} 
\noindent where prime stands for derivative with respect to $\mu$. The first fundamental form $\mathcal {I}$ and the second fundamental form $\mathcal {II}$ of surface~(\ref{gsi3}) read (see the Appendix)
\begin{align}\label{gsi6}
 &
 \mathcal {I}=\begin{pmatrix}
    E & 0\\ 
   0 & G
  \end{pmatrix} = \eta
  \begin{pmatrix}
    h^{-1} & 0\\ 
    0 &  h
  \end{pmatrix} 
  \\
  &
 \mathcal {II}= \begin{pmatrix}
    L & 0\\ 
   0 & N
  \end{pmatrix} = 
   \eta^{-2}\begin{pmatrix}
    f(g{}^{''}f'-g'f{}^{''})& 0\\ 
    0 & f^2g'
  \end{pmatrix} 
  \nonumber
\end{align} 
\noindent So, the Gaussian curvature $K_G$ for a metric of the above form is given by \cite{enc}
\begin{equation}\label{gsi7}
K_G=\frac{LN}{(EG)^2}=-\frac{1}{2}h{}^{''}\eta^{-2}=\eta^{-2}\left(1-\beta^2\left(1+3\mu^2\right)\right)\left(1-\beta^2\left(1-\mu^2\right)\right)^{-3}
\end{equation} 
\noindent It is useful to recall that the range of charge and total mass/energy for the family of Kerr-Newman black holes associated with the given parameters $\beta$ and $\eta$ are \cite{smarr2}:
\begin{align}\label{gsi8}
&0\leq Q\leq \eta\left(1-2\beta^2\right)^{1/2}\\
&\frac{1}{2}\eta\left(1-\beta^2\right)^{-1/2}\leq m\leq\eta\left(1-\beta\right)^2\nonumber
\end{align} 

\subsection{Case (i)} 

\noindent In this particular case, we find that the value of the \textit{distortion parameter} $\beta$ is $1/2$. We know that for this particular value of $\beta$ the Kerr metrics break into two classes separated by $\beta=1/2$ \cite{smarr2}. For $\beta<1/2$ the horizon black holes are oblately deformed spheres embedded in the Euclidean 3-space. They have everywhere positive Gaussian curvature. For $\beta>1/2$ there are regions of negative Gaussian curvature both on and around the axis of symmetry. In this case, the surface most resembles a hybrid sphere and pseudo-sphere. When $\beta=1/2$ (i.e., $a=m_{ir}$), at the poles  ($\mu=\pm 1$) the curvature of the event horizon surface becomes flat as the Gaussian curvature $K_G$ vanishes at these points. It is easily checked that for an uncharged rotation Kerr black hole, this happens when $a=\sqrt{3}/2\ \!m$. Beyond the value $\beta=1/2$, a global embedding in Euclidean 3-space is impossible. To sum up, \textit {the black hole corresponding to configuration (\ref{c6}) is an extreme Kerr-Newman black hole where the curvature at the poles becomes flat}.

\subsection{Case (ii)}

\noindent For the black hole~(\ref{c2}) we find 
\begin{equation}\label{gsi5}
\beta=\frac{1}{\sqrt{2}}(-\phi_-)=0.437016
\end{equation} 
\noindent Since the value of $\beta$ is less than $1/2$, the black hole horizon~(\ref{c2}) is a surface of revolution that can be embedded in Euclidean 3-space \cite{smarr2}. To gain more insight into the true intrinsic nature of the surface one must look locally. This task will be accomplished in the following subsections.

\subsubsection{Umbilic points}

\begin{figure}
\includegraphics[width=3cm]{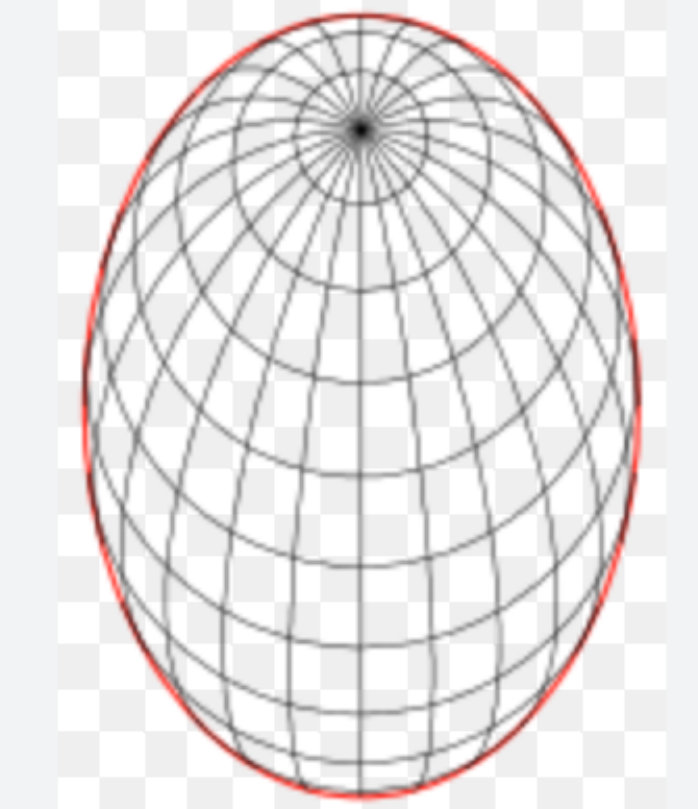}
\caption {\textit{Umbilic points. For metric~(\ref{gsi2}), the umbilic points are located at the poles.}}
\label{fig_EBH4}
\end{figure}

\noindent An \textit{umbilic} is a point on a surface where all normal curvatures are equal in all directions, and hence principal directions are indeterminate. Thus the orthogonal net of lines of curvature becomes singular at an umbilic. Actually, spheres and planes are the only surfaces whose points are umbilical. A non-flat umbilic occurs at an elliptic point where the principal forms $\mathcal{I}$ and $\mathcal{II}$ are proportional. So, at the umbilic we have the relation \cite{pressley}, \cite{hitchin}
\begin{align}\label{u1}
&\L=\kappa E\\
&G=\kappa N\nonumber\\
&F=\kappa M\qquad {\rm with}\qquad \kappa=constant \nonumber
\end{align}
\noindent It is easy to check that for metric~(\ref{gsi2}) we have $F=M=0$. In the Appendix, we show that for metric~(\ref{gsi2}) the umbilic should satisfy the equation
\begin{equation}\label{u2}
(1-\mu^2)\left(1-\beta^2\left(1+3\mu^2\right)\right)-\left(1-\beta^2\left(1-\mu^2\right)\right)^{4}+\mu^2=0
\end{equation}
\noindent When the limit $\beta=1/\sqrt{2}$ is exceeded a naked singularity occurs \cite{smarr2}. We can check that Eq.~(\ref{u2}) admits only one solution in the range $0\leq\beta\leq 1/\sqrt{2}$ corresponding to $\mu=\pm 1$. This solution is satisfied for all values of parameter $\beta$. In other words, \textit{for metric~(\ref{gsi2}) the umbilic points are located at the poles} (see Fig.~\ref{fig_EBH4}).

\subsubsection{The Pythagorean fundamental forms relation}
\noindent In the Appendix, we prove the following result. Let us consider the Kerr-Newman black hole surface with metric~(\ref{gsi2}) immersed into $E^3$ and satisfying at the umbilics the following \textit{Pythagorean fundamental forms relation} 
\begin{equation}\label{p1}
{\mathcal{I}}^2 + {\mathcal{II}}^2 = {\mathcal{III}}^2
\end{equation}
\noindent with $\mathcal{I}$, $\mathcal{II}$ and $\mathcal{III}$ denoting the matrices corresponding to the first, second, and third fundamental forms of the surface, respectively. Then, in the Planck unit system, at the umbilic points, the \textit{mean curvature of the surface} $H$ and the \textit{Gauss curvature of the surface} $K_G$ are respectively given by
\begin{align}\label{p2}
&H=(\phi_+)^{1/2}=(-\phi_-)^{-1/2}\\
&K_G=\phi_+=(-\phi_-)^{-1}\nonumber
\end{align}
\noindent This result is in line with that shown in \cite{aydin}. Furthermore, the value of the \textit{scale parameter} $\eta$ is 
\begin{equation}\label{p3}
\eta=-\phi_-
\end{equation}
\noindent We recall that the third fundamental form of the surface is the square of the differential of the unit normal vector surface at the point of a surface. The linear dependence relates the three fundamental forms
\begin{equation}\label{p4}
K_G\cdot {\mathcal{I}}-2H\cdot {\mathcal{II}}+{\mathcal{III}}=0
\end{equation}
\noindent To sum up, the main result of our analysis is: 

\noindent \textit{The extreme Kerr-Newman black hole}
\begin{align}\label{p5}
&{\bar m}=(-\phi_-)^{1/2}\\
&|Q|=(-\phi_-)^{3/2}\nonumber\\
&|{\bar L}|=(-\phi_-)^{5/2}\nonumber
\end{align}
\noindent \textit{satisfying the Pythagorean fundamental forms relation at the umbilic points, is the unique black hole where the \textit{scale parameter} $\eta$ is equal to a golden ratio} \footnote{Recall that we have adopted Planck's unit system.}. We recall that a sphere of surface area $4\pi r_G^2$ has $K_ G= 1/r_G^2$, which is constant and positive. For this particular black hole,  at the pole, the Gaussian curvature is (in Planck's unit system)
\begin{equation}\label{p4}
K_G=\frac{1}{r_G^2}\qquad where\qquad r_G=(-\phi_-)^{1/2}
\end{equation}
\noindent Hence, \textit{the surface of the extreme Kerr-Newman black hole~(\ref{c2}) satisfying the Pythagorean fundamental forms relation at the umbilic points is locally metrically a 2-sphere at these points with the radius of curvature equal to the square root of the golden ratio.}

\section{Removal of Energy from an Extremal Kerr-Newman Black Hole by Reversible Transformations}\label{er}
\noindent We are now interested in calculating the energy extraction from the extreme black hole (\ref{s9}). As seen, the extreme Kerr-Newman geometry corresponds to 
\begin{equation}\label{rs1}
m^2=Q^2+a^2
\end{equation}
\noindent From Eqs~(\ref{s1}) and (\ref{s3}) we get
\begin{align}\label{rs2}
&r_+=m\\
&m_{ir}=\frac{1}{2}\sqrt{m^2+a^2}\nonumber
\end{align}
\noindent The mass/energy formula~(\ref{s3}) tells us that the energy of a Kerr-Newman black hole is given by three contributions \cite{wheeler}: 1) the \textit{irreducible contribution to energy}, 2) the \textit{electromagnetic contribution to the energy}, and 3) the \textit{rotational contribution to the energy}. A black-hole transformation that holds fixed the irreducible mass is \textit{reversible}; one that increases it is \textit{irreversible}. The removable energy from a black hole $m_{extr}$ is clearly given by
\begin{equation}\label{rs3}
m_{extr}=m_{BH}^{(i)}-m_{BH}^{(f)}
\end{equation}
\noindent with $m_{BH}^{(i)}$ and $m_{BH}^{(f)}$ denoting the black hole energy before and after the extraction, respectively. By removing all of its electromagnetic and rotational energy from the black hole through a reversible transformation (i.e., $m_{ir}=const.$), we get $m_{BH}^{(f)}=m_{ir}$. Hence,
\begin{equation}\label{rs3a}
m_{extr}=m-m_{ir}
\end{equation}
\noindent with $m=m_{BH}^{(i)}$ given by Eq.~(\ref{rs1}). Our task is to determine the limit values of $m_{extr}$ i.e., $m_{extr}|_{min}$ and $m_{extr}|_{Max}$, for extreme Kerr-Newman black holes. Taking into account that $0\leq\beta\leq 1/{\sqrt{2}}$, or $0\leq a\leq {\sqrt{2}} m_{ir}$, from Eq.~(\ref{rs2}) we obtain that the energy that can be extracted from an extreme Kerr-Newman black hole by a \textit{reversible transformation} (i.e., $m_{ir}=const.$) is in the range of values
\begin{equation}\label{rs4}
\left(1-\frac{1}{\sqrt{2}}\right)m\leq m_{extr}<\frac{1}{2}m\qquad {\rm i.e.,}\qquad 29.28\% \leq \frac{m_{extr}}{m}< 50\%
\end{equation}
\noindent As previously mentioned, there are two geometrically distinct classes of Kerr-Newman black holes:

\noindent Class a): $0\leq\beta\leq 1/2$\qquad\ \!\qquad{\rm or}\qquad $0\leq a\leq m_{ir}$

\noindent Class b): $1/2 < \beta\leq 1/\sqrt{2}$ \qquad{\rm or}\qquad $m_{ir} < a \leq \sqrt{2} m_{ir}$

\subsection{Class a)}

\noindent This class consists of black holes with positive Gaussian curvature. These black holes are oblately deformed spheres and they can be embedded in $E^3$. So, the range of extractable energy from this class of extreme Kerr-Newman black holes is
\begin{equation}\label{rs4a}
\left(1-\frac{1}{\sqrt{3}}\right)m\leq m_{extr}<\frac{1}{2}m\qquad {\rm i.e.,}\qquad 42.26\% \leq \frac{m_{extr}}{m}< 50\%
\end{equation}

\subsection{Class b)}
\noindent This class consists of black holes where there are regions of negative Gaussian curvature both on and around the axis of symmetry. The surface of these black holes resembles a hybrid sphere and pseudo-sphere. This class is non-empty only if the charge is small enough and such that $\beta>1/2$ and $Q<\eta/\sqrt{2}$ \cite{smarr2}.  In this case, the range of extractable energy from this class of extreme Kerr-Newman black holes is
\begin{equation}\label{rs4b}
\left(1-\frac{1}{\sqrt{2}}\right)m\leq m_{extr}<\left(1-\frac{1}{\sqrt{3}}\right)m\qquad {\rm i.e.,}\qquad 29.28\% \leq \frac{m_{extr}}{m}< 42.26\%
\end{equation}

\subsection{Energy extractable from the extreme black holes~(\ref{c6}) and (\ref{c2})}
\noindent The energy we can extract from the black hole~(\ref{c6}) by reversible transformations amounts to 
\begin{equation}\label{rs5}
\frac{E_{extr}}{E_m}=\left(1-\frac{1}{\sqrt{3}}\right)=42.26\%
\end{equation}
\noindent while the energy extractable from the black hole~(\ref{c2}) by reversible transformations amounts to
\begin{equation}\label{rs6}
\frac{E_{extr}}{E_m}=\left(1-\frac{(-\phi_-)^{1/2}}{\sqrt{2}}\right)=44.41\%
\end{equation}
\noindent In other words, \textit{by reversible transformations, from the black hole~(\ref{c6}) it is possible to extract only the minimum possible energy. Furthermore, in percentage, the energy extractable from the black hole~(\ref{c2}) is greater than that of the black hole~(\ref{c6}), and amounts to $88.82\%$ of the maximum value extractable from an extreme black hole of Kerr-Newman}.

\section{Conclusions}\label{c}
We analyzed extreme black holes in the Kerr-Newman geometry. By looking at the geometrical version of the equations of mass/energy for extreme Kerr-Newman black holes, we found two meaningful cases. The first black hole~(\ref{c6}) corresponds to the case where the distortion parameter $\beta$ has the limit value $\beta=1/2$.  Alongside this case, there is another even more surprising one: there exists a particular extreme black hole~(\ref{c2}) where all its fundamental physical quantities (i.e., its mass, its charge, its angular momentum, and the extractible energy from the black hole by reversible transformations) are incommensurable with the black hole's irreducible mass. All these constants are expressed in terms of the golden ratio numbers $-\phi_-$. These extreme black holes have been studied in the Christodoulou diagram and their topology has been investigated with the tools of differential geometry. The first black hole~(\ref{c6}) corresponds to the limit case where the event horizon surface is flat at the poles. As to the second extreme black hole~(\ref{c2}), we prove the beautiful result that if this extreme black hole satisfies the Pythagorean fundamental forms relation at the umbilic points, then both the \textit{scale parameter} and the Gauss curvature of the surface at the pole are equal to the golden ratio numbers. Furthermore, at the poles, the mean curvature of the surface is equal to the square root of the golden ratio. In this case, the \textit{scaled fundamental quantities} mass, charge, and angular momentum are equal to the square root of the golden section raised to odd integers. Successively, we computed the energy extractible from these extreme black holes by reversible transformations. We showed that, by keeping the irreducible mass constant to its initial value, the energy we can extract from the extreme black hole~(\ref{c2}) amounts to $88.82\%$ of the maximum value that can be extracted from an extreme Kerr-Newman black hole. In percentage, the energy extractible from this extreme black hole is greater than the energy extractible from the black hole~(\ref{c6}). It is the authors' opinion that a clearer understanding of the surface geometry of the extreme Kerr-Newman family of black holes will shed light on other problems in the physics of black holes and, more generally, allow insight into astrophysical processes. Now, an open question is whether the geometric properties found for these extreme black holes have physical consequences. An analysis of this problem is now underway.

\acknowledgments
\noindent This work was carried out in collaboration with Prof. Remo Ruffini of ICRANet. The manuscript will soon be submitted for publication in international journals in alphabetical order (R. Ruffini and G. Sonnino).

\noindent GS is grateful to Prof. Pasquale Nardone of the Universit{\'e} Libre de Bruxelles (ULB) for useful discussions.

\appendix

\section{Geometrical interpretation of the two extremal black holes}

\begin{figure}
\includegraphics[width=4cm]{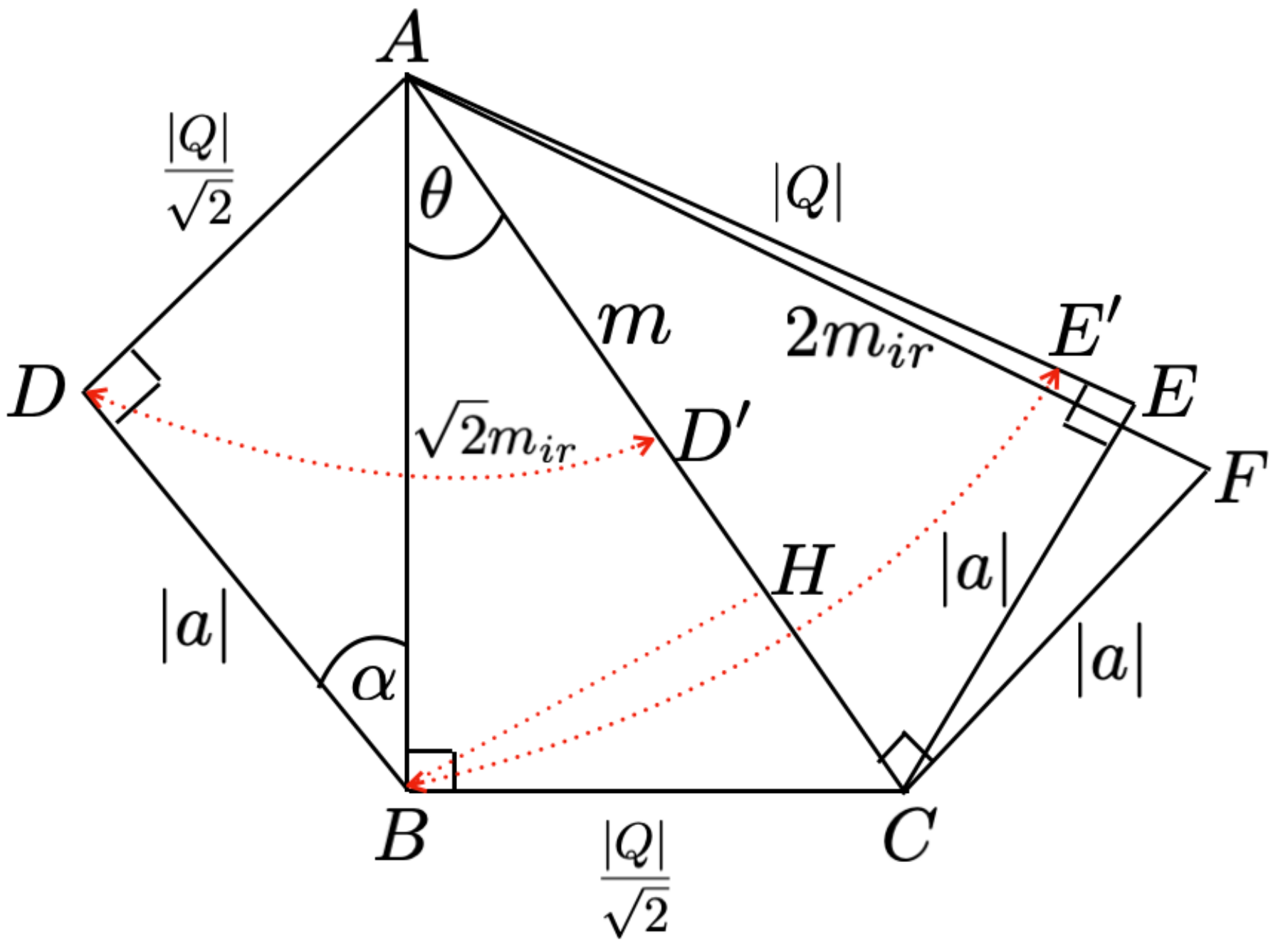}
\caption {\textit{Geometrical version of the extreme Kerr-Newman black holes equations. The right triangle ABC and the right triangle ADB are the first and the second equation of system~(\ref{s9}), respectively. Triangles ACE and ACF correspond to Eq.~(\ref{s10}) and Eq.~(\ref{s11}), respectively. Note that extreme Kerr-Newman black holes satisfy the relation $\sqrt{2}m_{ir}\sin\alpha=m\sin\theta$}.}
\label{fig_EBH1}
\end{figure}

\noindent Eqs~(\ref{s9}), (\ref{s10}), and (\ref{s11}) are equivalent from a geometrical point of view to four right triangles interconnected each with the other. This can easily be checked by looking at Fig.~\ref{fig_EBH1}.

\noindent From Fig.~\ref{fig_EBH1} we see that the two special cases (i) and (ii) investigated by us can be obtained by performing the rotations indicated with the red dotted arrows. More specifically,

\noindent Case (i)

\noindent This corresponds to the case where triangle ABC is exactly the same as triangle ACE (see Fig.~EBH5.pdf).
\begin{figure}
\includegraphics[width=3cm]{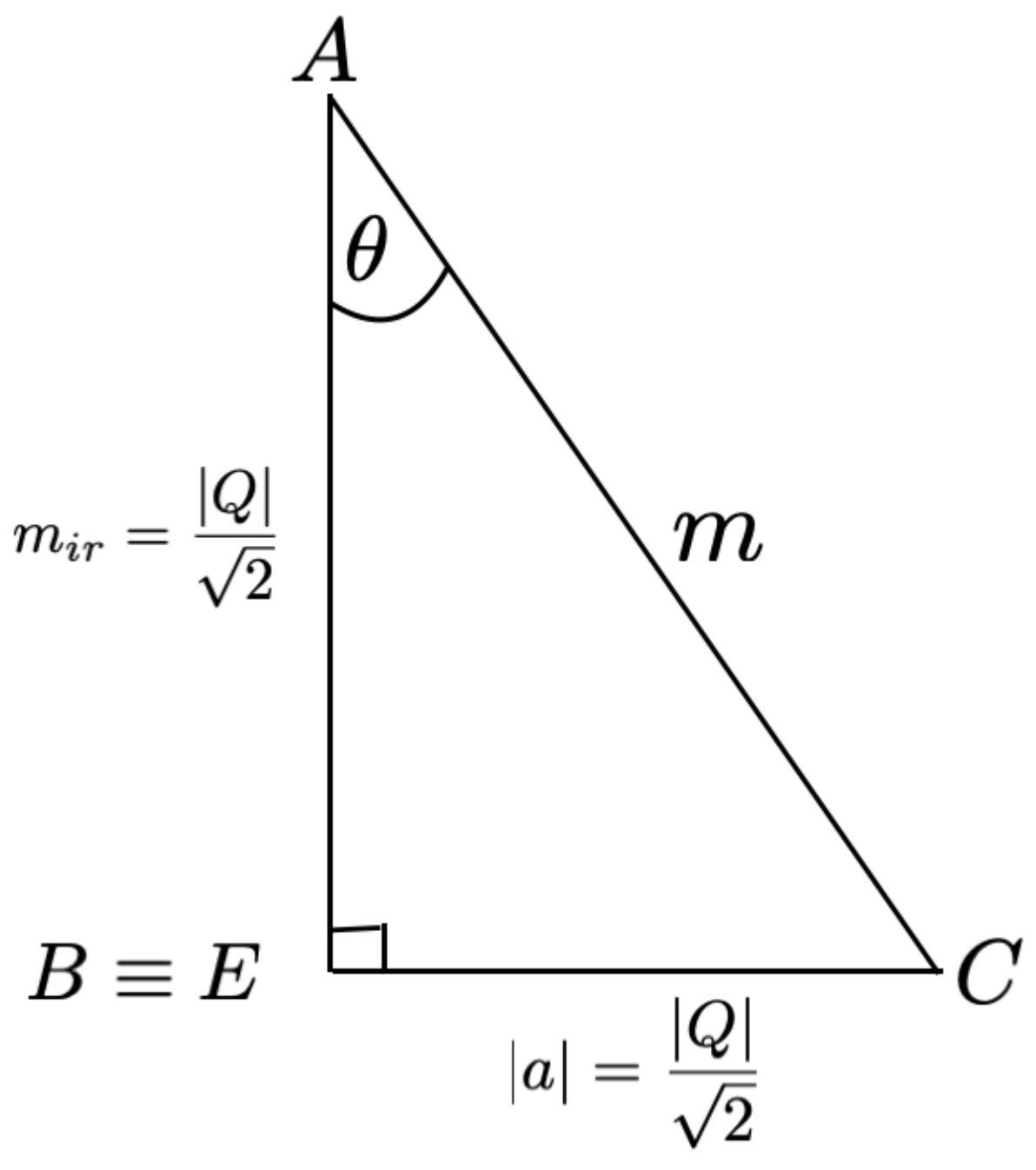}
\caption {\textit{Special configuration {\bf (i)}: triangle ABC is exactly the same as triangle AEC}.}
\label{fig_EBH5}
\end{figure}
\noindent Indeed, this condition implies:
\begin{equation}\label{Aa1}
|Q|=\frac{1}{\sqrt{2}}\eta
\end{equation}
\noindent So, from system~(\ref{s9}) we find
\begin{align}\label{Aa12}
&{\bar m}=\sqrt{\frac{3}{2}}\eta\\
&|Q|=\frac{1}{\sqrt{2}}\eta\nonumber\\
&|{\bar L}|=\frac{\sqrt{3}}{2}\eta^2 \nonumber
\end{align}

\noindent Case (ii) 

\noindent This corresponds to the case where one of the catheters of the right triangle ADB coincides exactly with the height of the triangle ABC relative to the base AC (see Fig.~EBH2.pdf).
\begin{figure}
\includegraphics[width=3cm]{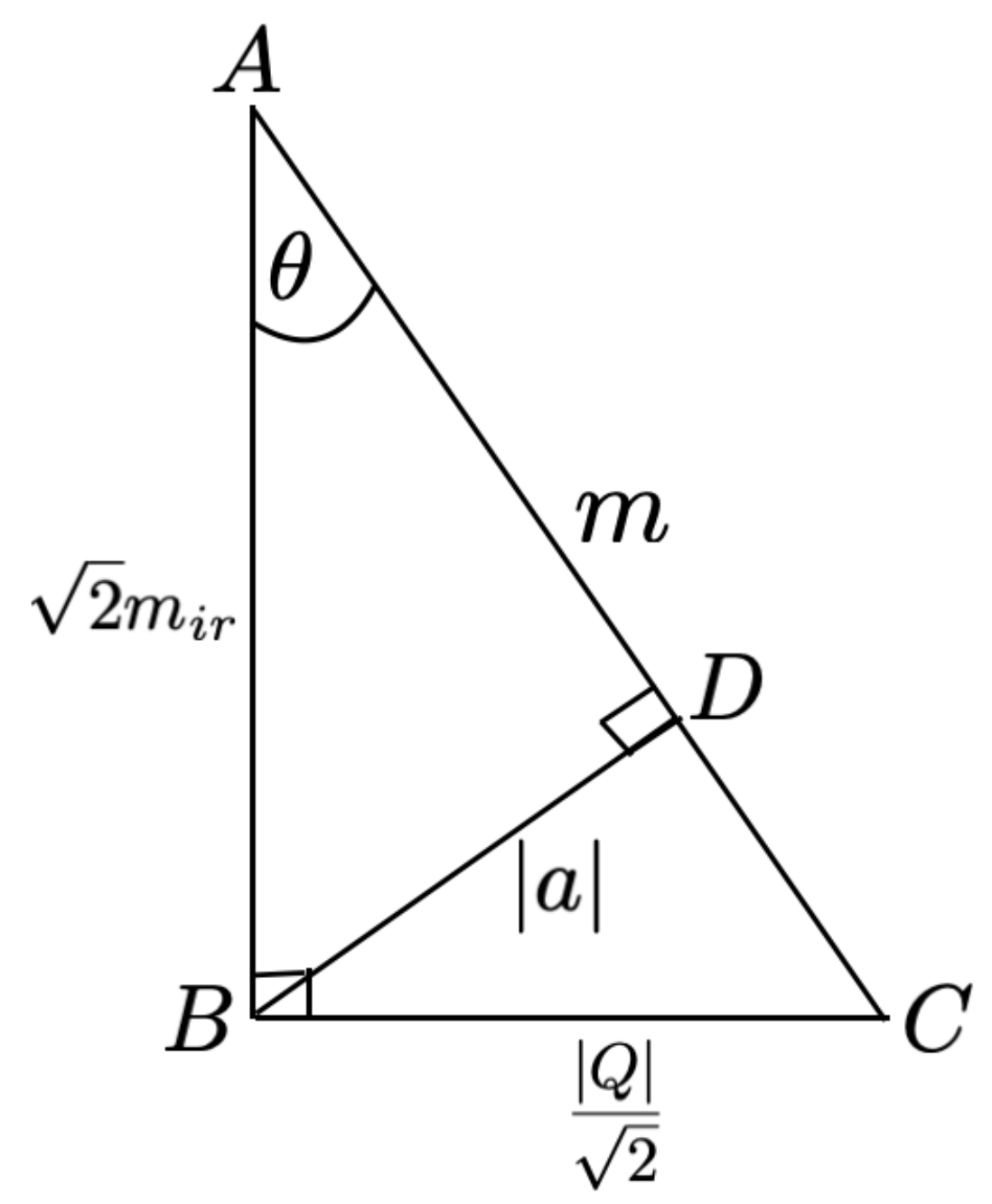}
\caption {\textit{Special configuration {\bf (ii)}: the catheter BD of the right triangle ADB coincides exactly with the height of the triangle ABC relative to the base AC}.}
\label{fig_EBH2}
\end{figure}
\noindent By imposing this condition for the catheter BD we get
\begin{equation}\label{Aa2}
{|\bar L|}=\eta |Q|
\end{equation}
\noindent which combined with Eqs~(\ref{s13}) gets
\begin{align}\label{Aa22}
&{\bar m}=(-\phi_-)^{-1/2}\eta\\
&|Q|=(-\phi_-)^{1/2}\eta\nonumber\\
&|{\bar L}|=(-\phi_-)^{1/2}\eta^2 \nonumber
\end{align}
\noindent where $\phi_-$ is one of the solutions of the golden ratio equation
\begin{equation}\label{c3}
\phi^2-\phi-1=0
\end{equation}
\noindent i.e.,
\begin{equation}\label{c4}
\phi_{-}= \frac{1-\sqrt{5}}{2}
\end{equation}
\section{Determination of the first and the second fundamental forms}
\noindent In these Appendices, we shall demonstrate the theorems used throughout this work. There are three types of so-called fundamental forms. The most important are the first and second (since the third can be expressed in terms of these). The first fundamental form (or line element) is given explicitly by the Riemannian metric \cite{weisstein}
\begin{equation}\label{a1}
ds^2=Edu^2+2Fdudv+Gdv^2
\end{equation} 
\noindent This form expresses the principal linear part of the growth of the arc length between two extremely close points located at the surface. The second fundamental form is given explicitly by
\begin{equation}\label{a2}
ds^2=\L du^2+2Mdudv+Ndv^2
\end{equation} 
\noindent The second fundamental form serves to define extrinsic invariants of the surface, i.e., its principal curvatures, and it is defined for a smooth immersed submanifold in a Riemannian manifold. If an arbitrary surface ${\mathcal{S}}$ is parametrized as
\begin{equation}\label{a3}
{\bf \mathcal{X}}=(x(u,v),y(u,v),z(u,v))
\end{equation} 
\noindent then the matrices corresponding to the first and the second fundamental forms ${\mathcal{I}}$ and ${\mathcal{II}}$ read, respectively
\begin{align}\label{a4}
& \mathcal {I}=\begin{pmatrix}
    E & F\\ 
   F & G
  \end{pmatrix} \\
  & \mathcal {II}=\begin{pmatrix}
    \L & M\\ 
   M & N
  \end{pmatrix} \nonumber
  \end{align}
\noindent where
\begin{align}\label{a5}
& E=({\bf \mathcal{X}}_u\cdot {\bf \mathcal{X}}_u);\quad F=({\bf \mathcal{X}}_u\cdot {\bf \mathcal{X}}_v);\quad G=({\bf \mathcal{X}}_v\cdot {\bf \mathcal{X}}_v)\\
& \L=({\bf \mathcal{X}}_{uu}\cdot {\bf \mathcal{N}});\quad M=({\bf \mathcal{X}}_{uv}\cdot {\bf \mathcal{N}});\quad N=({\bf \mathcal{X}}_{vv}\cdot {\bf \mathcal{N}})\quad{\rm with}\nonumber\\
&{\bf \mathcal{N}}=\frac{\epsilon({\bf \mathcal{X}}_u\wedge {\bf \mathcal{X}}_v)}{|{\bf \mathcal{X}}_u\wedge {\bf \mathcal{X}}_v|}\quad{\rm and}\quad W=(EG-F^2)^{1/2}
\nonumber
\end{align}
\noindent In Eq.~(\ref{a5}) the subscript at the bottom of the variable stands for \textit{derivative of the variable with respect to that subscript}. Furthermore, $\epsilon= +1$ if the triple of vectors $\{ {\bf \mathcal{X}}_u, {\bf \mathcal{X}}_v, {\bf \mathcal{N}}\}$ has a right-hand orientation and $\epsilon= -1$ in the opposite case. In the case of metric~(\ref{gsi2}), for an arbitrary surface of revolution in Euclidean 3-space ($E^3$) parametrized as
\begin{equation}\label{a6}
{\bf \mathcal{X}}=(f(\nu)\cos\theta,f(\nu)\sin\theta,g(\nu))
\end{equation} 
\noindent we get
\begin{align}\label{a6}
 &
 \mathcal {I}= \eta^2
  \begin{pmatrix}
    h^{-1} & 0\\ 
    0 &  h
  \end{pmatrix} 
  \\
  &
 \mathcal {II} = 
   \eta^{-2}\begin{pmatrix}
    f(g{}^{''}f'-g'f{}^{''})& 0\\ 
    0 & f^2g'
  \end{pmatrix} 
  \nonumber
\end{align}
\noindent The determinant of the ratio of the second with respect to the first one is the Gaussian curvature of the surface at the point \cite{enc}:
\begin{equation}\label{a7}
K_G=\frac{\L N-M^2}{W^2}
\end{equation} 
\noindent This quantity measures the geometry intrinsic to the horizon itself and is independent of the embedding space. The trace of this ratio,
\begin{equation}\label{a8}
H=\frac{EN+G\L-2FM}{2W^2}
\end{equation} 
\noindent defines the mean curvature of the surface at the point \cite{enc}. The extrinsic curvature depends on the embedding space. As the parameter $\beta$ varies, the shape of the black hole deforms i.e., the extrinsic curvature varies while preserving the intrinsic geometry. In the case of metric~(\ref{gsi2}), we get
\begin{align}\label{a8}
&K_G=\eta^{-2}\left(1-\beta^2(1+3\mu^2)\right)\left(1-\beta^2(1-\mu^2)\right)^{-3}\\
&H=\frac{1}{2}\eta^{-1} h^{-1/2}\left(1-\mu^2(1-\mu^2)^{-4}h^4\right)^{1/2}\left(1-\mu^2(1-\mu^2)^{-4}h^4+h_1\right)\quad{\rm with}\nonumber\\
&h=(1-\mu^2)\left(1-\beta^2(1-\mu^2)\right)^{-1}\nonumber\\
& h_1=\left(1-\beta^2(1+3\mu^2)\right)(1-\mu^2)\left(1-\beta^2(1-\mu^2)\right)^{-4}\nonumber
\nonumber 
\end{align}
\noindent The surface of a Kerr-Newman black hole can be globally embedded in $E^3$ if $\beta <1/2$ and cannot be globally embedded in $E^3$ if $\beta > 1/2$ \cite{smarr2}.

\section{Derivation of the Equation for the Umbilic Points}
\noindent An umbilic point is a point on a surface at which the curvature is the same in any direction. In the differential geometry of surfaces in three dimensions, umbilics are points on a surface that are locally spherical. At an umbilic, the first fundamental form $\mathcal{I}$ is proportional to the second fundamental form $\mathcal{II}$ \cite{enc1}. Hence,
\begin{equation}\label{b1}
\L=\kappa E\quad{\rm and}\qquad N=\kappa G\qquad{\rm with}\quad \kappa=constant
\end{equation} 
\noindent From Eq.~(\ref{b1}), we have
\begin{equation}\label{b2}
\frac{\L}{N}=\frac{E}{G}
\end{equation} 
\noindent From Eq.~(\ref{a6}) we have
\begin{equation}\label{b3}
\frac{\L}{N}=h^{-2}\qquad {\rm or}\qquad g^{''}f'-f^{''}g'=fg'(1-\mu^2)^{-2}\left(1-\beta^2(1-\mu^2)\right)^{2}
\end{equation} 
\noindent After simple algebra, we finally get the equation for umbilics
\begin{equation}\label{b4}
(1-\mu^2)\left(1-\beta^2\left(1+3\mu^2\right)\right)-\left(1-\beta^2\left(1-\mu^2\right)\right)^{4}+\mu^2=0
\end{equation}
\noindent 

\section {Solution of the Pythagorean fundamental forms equation}

\noindent In differential geometry, a surface immersed into 3-dimensional space forms is said to fulfill the \textit{Pythagorean-like formula} if the following relation among its fundamental forms is satisfied 
\begin{equation}\label{pf1}
\mathcal{I}^2 + \mathcal{II}^2 = \mathcal{III}^2
\end{equation}
\noindent with $\mathcal{I}$, $\mathcal{II}$, and $\mathcal{III}$ denoting the matrices corresponding to the surface's first, second, and third fundamental forms, respectively \cite{aydin1}. In differential geometry, the third fundamental form is a surface metric where, unlike the second fundamental form, it is independent of the surface normal. The third fundamental form is given in terms of the first and second forms by \cite{chase}
\begin{equation}\label{pf2}
K_G\cdot {\mathcal{I}}-2H\cdot {\mathcal{II}}+{\mathcal{III}}=0
\end{equation}
\noindent At the umbilic points, we have $\L=\kappa E$ and $N=\kappa G$. Hence, at the umbilics, the Gauss curvature of the surface and the mean curvature of the surface simply read
\begin{align}\label{pf3}
&K_G=\frac{\L N-M^2}{W^2}=\frac{\kappa^2 EG}{EG}=\kappa^2\\
&H=\frac{\L G+NE-2FM}{2 W^2}=\frac{2\kappa EG}{2 EG}=\kappa\nonumber
\end{align}
\noindent At the umbilics, the third fundamental form reads
\begin{equation}\label{pf4}
\mathcal{III}\equiv
\begin{pmatrix}
    P & 0\\ 
   0 & Q
  \end{pmatrix}
  =2H 
  \begin{pmatrix}
    \L & 0\\ 
   0 & N
  \end{pmatrix}
  -K_G
  \begin{pmatrix}
    E & 0\\ 
   0 & G
  \end{pmatrix}
\end{equation}
\noindent or
\begin{equation}\label{pf5}
\mathcal{III}\equiv
\begin{pmatrix}
    P & 0\\ 
   0 & Q
  \end{pmatrix}
  =2\kappa 
  \begin{pmatrix}
    \kappa E & 0\\ 
   0 & \kappa G
  \end{pmatrix}
  -\kappa^2
  \begin{pmatrix}
    E & 0\\ 
   0 & G
  \end{pmatrix}
  =\kappa^2
  \begin{pmatrix}
    E & 0\\ 
   0 & G
  \end{pmatrix}
\end{equation}
\noindent The horizon of a black hole satisfies the Pythagorean-like formula at the umbilics if
\begin{equation}\label{pf5}
\kappa^4
  \begin{pmatrix}
    E & 0\\ 
   0 & G
  \end{pmatrix}
  =
  \kappa^2
  \begin{pmatrix}
    E & 0\\ 
   0 & G
  \end{pmatrix}
  +
  \begin{pmatrix}
    E & 0\\ 
   0 & G
  \end{pmatrix}
\end{equation}
\noindent Eq.~(\ref{pf5}) is satisfied for all values of E and G if, and only if, the following equation is satisfied
\begin{equation}\label{pf6}
\kappa^4-\kappa^2-1=0
\end{equation}
\noindent The positive solution of Eq.~(\ref{pf6}) reads
\begin{equation}\label{pf7}
\kappa^2=\phi_+
\end{equation}
\noindent Hence,
\begin{equation}\label{pf8}
K_G=\phi_+\qquad{\rm and}\qquad H=\phi_+^{1/2}
\end{equation}
\noindent The main result of our analysis is: \textit{if the extreme Kerr-Newman black hole~(\ref{c2}) satisfies the Pythagorean-like formula at the umbilic points, then the horizon is locally metrically a 2-sphere at these points with the Gauss curvature of the surface equal to the golden ratio and the mean curvature of the surface equal to the square root of the golden ratio.}



\end{document}